\documentclass[twocolumn,secnumarabic,amssymb,nobibnotes,aps,superscriptaddress,pra]{revtex4-1}

\usepackage[dvips]{graphicx}
\usepackage{bm}
\usepackage{latexsym} 
\usepackage{amsmath}
\usepackage{amssymb}
\usepackage{color}

\newcommand{\ui}{{\rm i}}
\newcommand{\veps}{{\varepsilon}}
\newcommand{\eps}{{\epsilon}}

\newcommand{\bmr}{{\bm r}}
\newcommand{\bmp}{{\bm p}} 
\newcommand{\bmk}{{\bm k}}
\newcommand{\bmq}{{\bm q}}
\newcommand{\bra}{\langle}
\newcommand{\ket}{\rangle}
\newcommand{\kB}{k_{\rm B}}

\begin{document}

\title{
  Spin pumping into superconductors:\\
  A new probe of spin dynamics in a superconducting thin film 
}

\author{Masashi Inoue}
\affiliation{Department of Physics, Okayama University, Okayama 700-8530, Japan}
\author{Masanori Ichioka}
\affiliation{Research Institute for Interdisciplinary Science, Okayama University, Okayama 700-8530, Japan}
\author{Hiroto Adachi}
\affiliation{Research Institute for Interdisciplinary Science, Okayama University, Okayama 700-8530, Japan}

\date{\today}

\begin{abstract}
  Spin pumping refers to the microwave-driven spin current injection from a ferromagnet into the adjacent target material. We theoretically investigate the spin pumping into superconductors by fully taking account of impurity spin-orbit scattering that is indispensable to describe diffusive spin transport with finite spin diffusion length. We calculate temperature dependence of the spin pumping signal and show that a pronounced coherence peak appears immediately below the superconducting transition temperature $T_{\rm c}$, which survives even in the presence of the spin-orbit scattering. The phenomenon provides us with a new way of studying the dynamic spin susceptibility in a superconducting thin film. This is contrasted with the nuclear magnetic resonance technique used to study a bulk superconductor. 
\end{abstract}



\maketitle

\section{Introduction \label{Sec:I}}
Investigation of the interplay of superconductivity and magnetism has a long history. The seminal theoretical work by Abrikosov and Gor'kov~\cite{Abrikosov-Gorkov} describing the effects of magnetic impurities on superconductivity initiated the research field of gapless superconductivity~\cite{Maki-review}. Besides, the theoretical prediction on the appearance of nonuniform superconductivity by paramagnetic depairing effects~\cite{Fulde-Ferrell,Larkin-Ovchinnikov} has still had a major impact on the current research of superconductor/ferromagnet junctions~\cite{Buzdin05}. In the context of spintronics, the superconducting tunneling experiment has been applied to the measurement of the spin polarization in a number of materials~\cite{Tedrow73,Soulen98}, and an electrical spin injection into a superconductor is currently under active investigation~\cite{Yamashita02,Yang10,Wakamura14,Ohnishi14}. Furthermore, the interplay of superconductivity and magnetism has been a subject of intense debate~\cite{Pfleiderer09}. 

In those investigations mentioned above, the magnetic degrees of freedom are assumed to be in thermal equilibrium, and the {\it nonequilibrium} dynamics of the magnetization does not play any active role. In recent years, however, this nonequilibrium magnetization dynamics in magnetic heterostructures has drawn great attention as a new means for the spin current generation, which is now known as spin pumping~\cite{Tserkovnyak02}. In this method, the nonequilibrium magnetization dynamics in a ferromagnet is driven by ferromagnetic resonance (FMR), and it gives rise to the spin injection into the adjacent target material by transferring spin angular momentum through the $s$-$d$ exchange interaction at the interface. Because the spin pumping relies only on the spin dynamics and thus enables a charge-free spin injection~\cite{Ando11}, by now it is widely used as a versatile spin injection method. Indeed, the target materials range from normal metals~\cite{Urban01,Mizukami02,Saitoh06,Mosendz10,Sandweg11,Czeschka11,Castel12,Hahn13}, semiconductors~\cite{Ando11,Ando12,Rojas13,Shikoh13}, magnetic metals~\cite{Hyde14,Mendes14,Saglam16} and insulators~\cite{Wang14}, to more exotic systems such as graphene~\cite{Patra12,Tang13}, organic materials~\cite{Ando13,Qiu15}, a topological insulator~\cite{Shiomi14}, and a Rashba system~\cite{Sanchez13}. Given this background, it is worth investigating and clarifying the nature of spin pumping into superconductors. 

Experimentally, the spin pumping into a superconducting material was studied almost a decade ago in a Ni$_{80}$Fe$_{20}$/Nb bilayer system~\cite{Bell08}, and a decrease in the spin pumping signal was observed below the superconducting transition temperature $T_{\rm c}$. On the theoretical side, one of the recent progresses is the theoretical finding that the spin pumping signal is intimately related to the dynamic spin susceptibility of target materials~\cite{Ohnuma14}, which was derived in a close analogy to the linear-response approach to the spin Seebeck effect~\cite{Adachi11,Adachi13}. Indeed, a number of experiments suggesting the correlation between the spin susceptibility and the spin pumping have been accumulating~\cite{Frangou16,Qiu16,Lin16}. Therefore, from the current theoretical point of view, the problem of calculating spin pumping into superconductors is reduced to the evaluation of the dynamic spin susceptibility in the superconducting state. 

The dynamic spin susceptibility in superconductors has long been studied using nuclear magnetic resonance (NMR)~\cite{MacLaughlin76}, especially by focusing on the behavior of nuclear-spin relaxation rate~\cite{Hebel-Slichter59,Masuda-Redfield62}. Then, a naive expectation would be that in the literature one could find a detailed calculation of the dynamic spin susceptibility in a superconducting state. In actual fact, however, existing theories discussing the NMR nuclear-spin relaxation rate focus only on a system without impurity spin-orbit scattering where the spin diffusion length diverges~\cite{Fulde68}, whereas we do not encounter such a situation even in a material with relatively long spin diffusion length, such as Al~\cite{Bass07}. Therefore, in order to obtain a physically reasonable result with nondiverging spin diffusion length, it is of crucial importance to deal with the impurity spin-orbit scattering in an adequate manner. As far as we know, however, no such theoretical calculation has been reported in the literature. 

In this paper we present a theory of spin pumping into superconducting materials, by fully taking account of the impurity spin-orbit scattering. Making use of the previous theoretical finding~\cite{Ohnuma14} that the strength of the spin pumping is proportional to the imaginary part of the dynamic spin susceptibility of target materials, we calculate temperature dependence of the spin pumping signal in a bilayer composed of an insulating ferromagnet and an $s$-wave superconductor. By evaluating the dynamic spin susceptibility of the superconducting target material, we show that a pronounced coherence peak appears in the signal immediately below the superconducting transition temperature $T_{\rm c}$, which survives even in the presence of the impurity spin-orbit scattering. Since existing NMR technique is not suitable for thin film samples, we further argue that the phenomenon under discussion can be used as a new method for detecting the spin dynamics in a superconducting thin film. 

The outline of the paper is as follows. In Sec.~\ref{Sec:II}, our model is given. In Sec.~\ref{Sec:III}, the formulation to derive the dynamic spin susceptibility of a superconductor is presented, by fully taking account of the vertex corrections due to impurity spin-orbit scattering. In Sec.~\ref{Sec:IV}, temperature dependence of the spin pumping into an $s$-wave superconductor is numerically evaluated for various strengths of the impurity spin-orbit scattering. Finally, in Sec.~\ref{Sec:V} we discuss and summarize our results. We use units $\hbar=\kB=1$ throughout this paper.

\begin{figure}[t] 
  \begin{center}
        \scalebox{0.38}[0.38]{\includegraphics{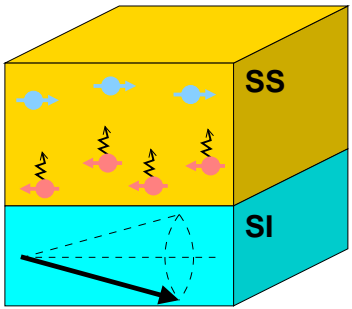}}
  \end{center}
\caption{ 
  (Color online) Schematic view of the system considered in this paper, where a spin sink (SS) is placed on top of a spin injector (SI). The SS is an $s$-wave superconductor in the present paper, whereas it was assumed to be a weak itinerant ferromagnet in Ref.~\cite{Ohnuma14}. In both cases, the SI is an insulating magnet. 
}
\label{fig1_Inoue}
\end{figure}

\section{Model \label{Sec:II}}
As in Ref.~\cite{Ohnuma14}, we consider a bilayer composed of a spin injector (SI) and a spin sink (SS), as shown in Fig.~\ref{fig1_Inoue}. In Ref.~\cite{Ohnuma14}, the SS was assumed to be a weak itinerant ferromagnet such as NiPd. In the present paper, by contrast, we consider a situation where the SS is an $s$-wave superconductor such as Nb. In both cases, the SI is a magnetic insulator, most typically yttrium iron garnet (YIG). We assume that a static magnetic field ${\bm H}_{0}= H_0 {\bm \hat{\bm z}}$ is applied to the bilayer in the lateral direction, and that the anisotropy field is much weaker than $H_0$ such that it can be disregarded. 

We focus on the situation where an external microwave with the angular frequency $\omega_{\rm ac}$ is applied to the SI/SS bilayer to drive the ferromagnetic resonance (FMR) of the SI. In the magnon language the external microwave resonantly excites uniform magnon mode inside the SI, with the angular frequency $\omega_{\rm ac}$. Without the attachment of the SS, the magnon has an intrinsic damping rate $\alpha_0 \omega_{\rm ac}$, where $\alpha_0$ is the (dimensionless) intrinsic Gilbert damping constant. In the presence of the SS, because there arises an additional spin dissipation channel, the magnon acquires an additional Gilbert damping so that the total Gilbert damping constant $\alpha$ is given by 
\begin{equation}
  \alpha= \alpha_0 + \delta \alpha, 
\end{equation}
where $\delta \alpha$ is the additional Gilbert damping constant. In FMR experiments, the additional Gilbert damping constant is obtained from the FMR linewidth $\Delta H$ through the relation
\begin{equation}
  \gamma \Delta H = \frac{2}{\sqrt{3}} (\alpha_0 + \delta \alpha) \omega_{\rm ac},
  \label{eq:Hpp01}
\end{equation}
where $\gamma$ is the gyromagnetic ratio. The appearance of the additional spin dissipation channel means that spins are pumped into the SS, and thus this phenomenon is termed spin pumping. Because the spins pumped into the SS diffuse in the form of a spin current, this phenomenon has drawn much attention as a new means of spin current generation.

\begin{figure}[t] 
  \begin{center}
    \scalebox{0.35}[0.35]{\includegraphics{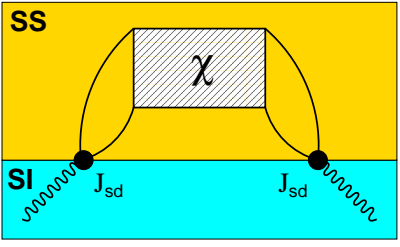}}
  \end{center}
\caption{ 
  (Color online) Diagrammatic representation of the magnon selfenergy leading to the additional Gilbert damping. A solid line is the electron Green's function, a wavy line the magnon propagator, $\chi$ the spin susceptibility, and a black dot the $s$-$d$ interaction at the SS/SI interface. 
}
\label{fig_selfenergy01}
\end{figure}

Now we briefly review the linear-response approach to the spin pumping that has been developed in Refs.~\cite{Ohnuma14,Ohnuma15}. As already mentioned, the additional Gilbert damping is regarded in the magnon language as the additional damping rate of the collective mode, i.e., the magnon. According to the framework of many-body theory~\cite{AGD}, the damping of a collective excitation can be calculated from the corresponding selfenergy $\Sigma^R(\omega)$. In the present case, the selfenergy is diagrammatically expressed in Fig.~\ref{fig_selfenergy01}, from which we see that it is related to the dynamic spin susceptibility $\chi^R_{\bm q}(\omega)$ of the SS as
\begin{equation}
  \Sigma^R(\omega)= -\frac{\bra \bra J^2_{\rm sd}\ket \ket}{\hbar^2}\sum_{\bm q} \chi_{\bm q}^R(\omega), 
\end{equation}
where $J_{\rm sd}$ is the $s$-$d$ interaction at the SS/SI interface, $\bra \bra J^2_{\rm sd} \ket \ket =2 J^2_{\rm sd} S_0 N_{\rm int}/( N_{\rm SI} N_{\rm SS})$ with $N_{\rm int}$, $N_{\rm SI}$, and $N_{\rm SS}$ being the number of localized spins at the interface, and the number of lattice sites in the SI and SS, respectively. Using the relation $\delta \alpha \, \omega_{\rm ac}= - {\rm Im} \Sigma^R(\omega_{\rm ac})$, we arrive at the final result: 
\begin{equation}
  \delta \alpha = \frac{\bra \bra J^2_{\rm sd}\ket \ket}{\hbar^2}
  \sum_{\bm q} \frac{1}{\omega_{\rm ac}} {\rm Im} \chi_{\bm q}^R(\omega_{\rm ac}). 
  \label{eq:Spump01}
\end{equation}
Therefore, the remaining task is to calculate the dynamic spin susceptibility in the superconducting state.

As mentioned in Sec.~\ref{Sec:I}, when calculating the dynamic spin susceptibility of superconductors, it is quite important to take account of the spin-orbit scattering by impurities in a proper way, because it is indispensable to describe the spin dissipation that produces a finite spin diffusion length. As for the {\it static} spin susceptibility in superconductors such a calculation has been known~\cite{Abrikosov62}, which reveals that the spin-orbit scattering by impurities completely modifies the behavior of the static susceptibility, and the resultant temperature dependence of the susceptibility deviates substantially from that of a pure superconductor represented by the Yoshida function~\cite{Yoshida58}. Regarding the {\it dynamic} spin susceptibility in the superconducting state, on the other hand, there has been no such calculation reported so far except for Ref.~\cite{Maki73}, where the analysis is limited only to a narrow gapless region in the immediate vicinity of the superconducting transition temperature $T_{\rm c}$ and in the presence of pair-breaking perturbation. Therefore, we adopt such a route that the method of Ref.~\cite{Abrikosov62} for the {\it static} spin susceptibility is extended to the {\it dynamic} spin susceptibility, by employing the procedure of analytic continuation from the Matsubara susceptibility to the retarded susceptibility.

We begin with the following Hamiltonian for the SS: 
\begin{eqnarray}
  {\cal H} &=& {\cal H}_{\rm BCS}+ {\cal H}_{\rm imp}, 
\end{eqnarray}
where the first term, 
\begin{eqnarray}
  {\cal H}_{\rm BCS} &=&  \sum_{\bmp} c^\dag_{\bmp} \xi_\bmp  c_{\bmp}
  - \frac{|g|}{V} \sum_{\bmp,{\bm p'}, \bmk} c^\dag_{\bmp+\bmk \uparrow} c^\dag_{-\bmp \downarrow} 
  c_{-\bmp'+\bmk \downarrow}c_{\bmp' \uparrow}, \nonumber \\
\end{eqnarray}
is the BCS Hamiltonian with attractive interaction $|g|$. Here, $c^\dag_{\bmp}= (c^\dag_{\bmp,\uparrow}, c^\dag_{\bmp,\downarrow})$ is the electron creation operator for spin projection $\uparrow$ and $\downarrow$, $\xi_\bmp$ is the kinetic energy measured from the chemical potential, and $V$ is the system volume. The second term, 
\begin{eqnarray}
  {\cal H}_{\rm imp} &=&
  \sum_{\bmp,\bmp'} c^\dag_{\bmp} \hat{U}_{\bmp,\bmp'} c_{\bmp'}, 
\end{eqnarray} 
describes the scattering by impurities, where
\begin{equation}
  \hat{U}_{\bmp,\bmp'}
  = u_{0}({\bmp-\bmp'})+ \ui u_{\rm so} ({\bmp-\bmp'})\hat{\bm \sigma} \cdot (\bmp \times \bmp')
\end{equation}
is the impurity potential with $\hat{\bm \sigma}$ being the Pauli matrices, and $u_{0}({\bmp-\bmp'})$ and $u_{\rm so}({\bmp-\bmp'})$ are the amplitude of the momentum scattering and the spin-orbit scattering, respectively. Note that in the present paper, we use a notation to represent an operator in the spin space by using ``hat'' such as $\hat{O}$.

Following Ref.~\cite{Gorkov64}, the Gor'kov equation for the present system is given by 
\begin{eqnarray}
  \Big[
    \check{L}_\bmp
    + \check{V}_{\bmp,\bmp'}
      \Big]
  \check{G}_{\bmp,\bmp'}(\ui \veps_n)
  = (2 \pi)^3 \delta(\bmp- \bmp') \check{1}, 
\end{eqnarray}
where
\begin{equation}
  \check{L}_\bmp =
  \begin{pmatrix}
    \ui \veps_n - \xi_\bmp, & \ui \hat{\sigma}_y \Delta\\
    \ui \hat{\sigma}_y \Delta, & - \ui \veps_n- \xi_\bmp
  \end{pmatrix}  
\end{equation}
defines the Green's function in the pure system, $\veps_n= 2 \pi T (n+1/2)$ is a fermionic Matsubara frequency with $n$ being an integer, $\Delta$ is the superconducting gap, and we denote a matrix in the particle-hole (Nambu) space by ``check'' accent. The effect of impurities is described by 
\begin{equation}
  \check{V}_{\bmp,\bmp'} =
  \begin{pmatrix}
    \hat{U}_{\bmp,\bmp'}, & 0 \\
    0, & \hat{U}^t_{\bmp',\bmp}
  \end{pmatrix}, 
\end{equation}
where $\hat{U}^t$ means the transpose of a matrix $\hat{U}$ in the spin space. As in Ref.~\cite{Gorkov64}, the impurity-averaged Green' function, 
\begin{equation}
  \check{G}_\bmp (\ui \veps_n) =   
  \begin{pmatrix}
    {\cal G}_\bmp (\ui \veps_n) , & {\cal F}_\bmp (\ui \veps_n) \ui \hat{\sigma}^y \\
    {\cal F}^\dag_\bmp (\ui \veps_n)\ui \hat{\sigma}^y , & {\cal G}^\dag_\bmp (\ui \veps_n) 
  \end{pmatrix}, 
\end{equation}
plays the role of the zeroth-order Green's function in the present approach. Using the selfconsistent Born approximation for the impurity potential, we obtain 
\begin{eqnarray}
  \check{G}_\bmp (\ui \veps_n) &=&
   \frac{-1}{\widetilde{\veps}_n^2+ \widetilde{\Delta}^2 +\xi_\bmp^2}
  \begin{pmatrix}
    (\ui \widetilde{\veps}_n+ \xi_\bmp) , & \widetilde{\Delta} \ui \hat{\sigma}_y \\
    \widetilde{\Delta} \ui \hat{\sigma}_y ,  & (-\ui \widetilde{\veps}_n+ \xi_\bmp) 
  \end{pmatrix}, 
  \label{eq:G0v01}
\end{eqnarray}
where the Matsubara frequency $\veps_n$ and the superconducting gap $\Delta$ have the common selfenergy corrections in the present situation~\cite{Abrikosov62}: 
\begin{eqnarray}
  \widetilde{\veps}_n &=& \veps_n \eta, \label{eq:eps_tilde01}\\
  \widetilde{\Delta} &=& \Delta \eta, \label{eq:Dlt_tilde01}\\
  \eta &=& 1+ \frac{\Gamma_{(+)}}{\sqrt{\veps_n^2+\Delta^2}} . \label{eq:eta01} 
\end{eqnarray}
In the above equation, the scattering rate $\Gamma_{(+)}$ has two contributions as 
\begin{eqnarray}
  \Gamma_{(+)} &=& \frac{1}{2}\left( \frac{1}{\tau_0}+ \frac{1}{\tau_{\rm so}} \right), 
\end{eqnarray}
where $\tau_0$ is the momentum relaxation time and $\tau_{\rm so}$ is the spin-orbit relaxation time, which are respectively given by
${1}/{\tau_0} = 2 \pi N(0) n_{\rm imp} |u_0(0)|^2$ and
${1}/{\tau_{\rm so}} = 2 \pi N(0) n_{\rm imp} |u_{\rm so}(0)|^2 \bra ({\bmp} \times {\bmp}')^2 \ket_{\rm FS}$. Here, $N(0)$ is the density of states of electrons at the Fermi level, $n_{\rm imp}$ the number density of impurities, and $\bra \cdots \ket_{\rm FS}$ means the average over the Fermi surface.

Finally, the superconducting gap is determined selfconsistently by the gap equation, 
\begin{equation}
  \Delta = \frac{|g|}{V} \sum_{\veps_n} \sum_{\bmp}{\cal F}_\bmp (\ui \veps_n). 
\end{equation}
Because of the common form of the selfenergy corrections given by Eqs.~(\ref{eq:eps_tilde01})-(\ref{eq:eta01}), the superconducting transition temperature $T_{\rm c}$ as well as the gap equation remains the same as in the pure case, i.e., 
\begin{eqnarray}
  \ln \left( \frac{T}{T_{\rm c}} \right) \Delta &=&
  2 \pi T \sum_{\veps_n>0} \left( \frac{\Delta}{\sqrt{\veps_n^2+ \Delta^2}} 
  - \frac{\Delta}{\veps_n} \right), 
\end{eqnarray}
where we used the relation $1/|g|= \ln(T/T_{\rm c})+ 2 \pi T \sum_{\veps_n>0} (1/\veps_n)$. This is a consequence of Anderson's theorem~\cite{Anderson59} because, unlike the case of magnetic impurity scattering, the momentum scattering as well as the impurity spin-orbit scattering preserves the time-reversal symmetry of the electron system.

In Appendix~\ref{appendix1}, we briefly review how the present formalism is applied to the calculation of the static susceptibility of a superconductor with sizable impurity spin-orbit scattering~\cite{Abrikosov62}: 
\begin{equation}
  \chi_0 = N(0)\left( 1 - \sum_{\veps_n}
  \frac{\pi T \Delta^2}{(\veps_n^2+ \Delta^2)}
    \frac{1}{\sqrt{\veps_n^2+\Delta^2}+ \frac{2}{3 \tau_{\rm so}} } \right).
  \label{eq:GeYoshida01}
\end{equation}
The above result is an extension of the Yoshida function~\cite{Yoshida58} to the case with impurity spin-orbit scattering, which leads both to a nondiverging spin diffusion length and to a finite susceptibility even at zero temperature.

\section{Dynamic spin susceptibility in superconducting states \label{Sec:III}}

\begin{figure}[t] 
  \begin{center}
        \scalebox{0.35}[0.35]{\includegraphics{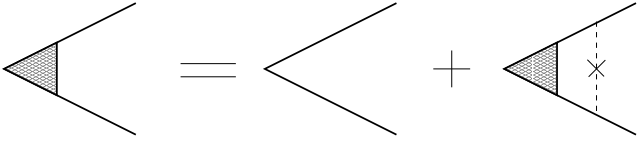}}
  \end{center}
  \caption{ 
    Diagrammatic representation of the vertex correction for $\delta \check{G}$. A dashed line with a cross means the scattering by impurities.}
  \label{fig_VC01}
\end{figure}

As explained in the previous section, temperature dependence of the additional Gilbert damping constant $\delta \alpha$ is obtained by calculating the dynamic spin susceptibility $\chi^R_\bmq(\omega_{\rm ac})$ [see Eq.~(\ref{eq:Spump01})]. Then, our strategy is to extend the method in Ref.~\cite{Abrikosov62} for the calculation of the static susceptibility $\chi_0$ into that of the dynamic susceptibility $\chi^R_\bmq(\omega_{\rm ac})$. The procedure consists of two steps. First, we generalize the {\it uniform} and {\it static} external magnetic field assumed in Ref.~\cite{Abrikosov62} into a space-time dependent external magnetic field, and consider the following external Hamiltonian 
\begin{eqnarray}
  \delta {\cal H}(\tau) &=& 
  - \int d^3 r \;
  {\bm \sigma} (\bmr,\tau) \cdot {\bm h} (\bmr,\tau),
  \label{eq:delH01}
\end{eqnarray}
where ${\bm h}(\bmr,\tau)= h_{\rm ac} \widehat{\bm z} \exp(-\ui \Omega_\nu \tau+ \ui \bmq \cdot \bmr)$. Here, the Bohr magneton $\mu_{\rm B}$ is absorbed into the definition of $h_{\rm ac}= \mu_{\rm B} H_{\rm ac}$, $\Omega_\nu= 2\pi T \nu$ is a bosonic Matsubara frequency with a positive integer $\nu>0$, and the time evolution is defined along the imaginary time $\tau$ as $c (\bmr,\tau)= e^{\tau {\cal H}} c(\bmr,\tau) e^{-\tau {\cal H}}$. Next, we calculate the resultant dynamic Matsubara susceptibility $\chi_\bmq(\ui \Omega_\nu)$, and perform an analytic continuation from the Matsubara susceptibility into the retarded susceptibility using the relation 
\begin{equation}
  \chi^R_\bmq(\omega_{\rm ac})= \chi_\bmq(\ui \Omega_\nu \to \omega_{\rm ac}+ \ui 0_+) , 
  \label{eq:continue01}
\end{equation}
which is detailed below.

In the presence of the space-time dependent external magnetic field, the matrix Green's function is, up to the linear order in $h_{\rm ac}$, written as 
\begin{equation}
  \check{G}_{\bmp,\bmq} (\ui \veps_n, \ui \Omega_\nu)
  = \check{G}_\bmp(\ui \veps_n)+ \delta \check{G}_{\bmp,\bmq}(\ui \veps_n, \ui \Omega_\nu) , 
  \label{eq:Gfull02}
\end{equation}
where $\veps_n$ and $\bmp$ are internal frequency and momentum whereas $\Omega_\nu$ and $\bmq$ are the external frequency and momentum. From the linear-response contribution to the Green's function $\delta \check{G}_{\bmp,\bmq}(\ui \veps_n, \ui \Omega_\nu)$, the Matsubara susceptibility is calculated to be 
\begin{equation}
  \chi_\bmq(\ui \Omega_\nu) = - \frac{\partial}{\partial h_{\rm ac}}  \frac{T}{2} \sum_{\veps_n} \int_\bmp \;
      {\rm Tr} \Big[ \hat{\sigma}^z \delta \hat{\cal G}_{\bmp, \bmq} (\ui \veps_n, \ui \Omega_\nu ) \Big],
      \label{eq:chiQW01}
\end{equation}
where $\delta \hat{\cal G}_{\bmp, \bmq} (\ui \veps_n, \ui \Omega_\nu )$ is the $(1,1)$ component of the matrix Green's function $\delta \check{G}_{\bmp,\bmq}(\ui \veps_n, \ui \Omega_\nu)$, and we have introduced the shorthand notation $\int_\bmp= \int \frac{d^3 p}{(2 \pi)^3}$. In the above equation, the factor $1/2$ arises because of the relation $2 \chi= \chi^{zz}$ in the paramagnetic phase, where $\chi^{zz}$ is the longitudinal susceptibility. The linear-response contribution to the Green's function can be expressed as 
\begin{equation}
  \delta \check{G}_{\bmp,\bmq}(\ui \veps_n,\ui \Omega_\nu) = \check{G}_{\bmp-\bmq}(\ui \veps_n- \ui \Omega_\nu)
  \check{\Lambda}_\bmq (\ui \veps_n,\ui \Omega_\nu) \check{G}_\bmp(\ui \veps_n). 
  \label{eq:delG_QW01}
\end{equation}
Here, the vertex function $\check{\Lambda}_\bmp(\ui \veps_n)$ representing the effects of impurity ladder shown in Fig.~\ref{fig_VC01} satisfies the following equation: 
\begin{eqnarray}
  \check{\Lambda}_{\bmq} (\ui \veps_n,\ui \Omega_\nu) &=&
  h_{\rm ac} \hat{\sigma}^z + n_{\rm imp} \int_{\bmp'} \check{V}_{\bmp,\bmp'}
  \check{G}_{\bmp'-\bmq} (\ui \veps_n- \ui \Omega_\nu)  \nonumber \\
  && \times \check{\Lambda}_\bmq (\ui \veps_n,\ui \Omega_\nu)
  \check{G}_{\bmp'}(\ui \veps_n) \check{V}_{\bmp',\bmp}.
  \label{eq:Lambda02}
\end{eqnarray}

\begin{figure}[t] 
  \begin{center}
        \scalebox{0.5}[0.5]{\includegraphics{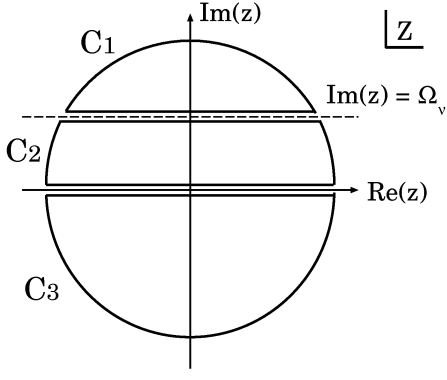}}
  \end{center}
  \caption{ 
    The contour to transform the Matsubara sum into an integral over $\eps={\rm Re}(z)$. 
  }
  \label{fig_contour01}
\end{figure}

Now we perform the analytic continuation from the Matsubara susceptibility into the retarded susceptibility [Eq.~(\ref{eq:continue01})], by using the formula
\begin{equation}
  T \sum_{\veps_n} g(\ui \veps_n) = \int_C \frac{d z }{4 \pi \ui } \tanh \left( \frac{z}{2T}\right) g(z), 
\end{equation}
where the contour is shown in Fig.~\ref{fig_contour01}. Following the standard procedure~\cite{AGD}, we find that the contour $C_1$ and $C_3$ result in the nondissipative part ${\rm Re} \chi^R_\bmq(\omega_{\rm ac})$ which is not of interest to us because of Eq.~(\ref{eq:Spump01}). By contrast, the contour $C_2$ gives us the dissipative part: 
\begin{eqnarray}
  {\rm Im} \chi^R_\bmq(\omega_{\rm ac}) &=& \frac{\partial}{\partial h_{\rm ac}}  \frac{1}{2}
  \int_{-\infty}^{\infty} \frac{d \eps}{4 \pi} \int_\bmp \;
  \left[ \tanh \left(\frac{\eps}{2T} \right) \right. \nonumber \\
    &-& \left. \tanh \left(\frac{\eps- \omega_{\rm ac}}{2T} \right) \right]
       {\rm Tr} \Big[ \hat{\sigma}^z \delta \hat{\cal G}^{RA}_{\bmp, \bmq} (\eps, \omega_{\rm ac}) \Big],
       \label{eq:chiQW02}
\end{eqnarray}
where $\delta \hat{\cal G}^{RA}$ is the $(1,1)$ component of 
\begin{equation}
  \delta \check{G}^{RA}_{\bmp,\bmq}(\eps,\omega_{\rm ac}) = \check{G}^R_{\bmp}(\eps)
  \check{\Lambda}^{RA}_\bmq (\eps,\omega_{\rm ac}) \check{G}^A_{\bmp-\bmq}(\eps-\omega_{\rm ac}). 
  \label{eq:delG_RA01}
\end{equation}
In the above equation, $\check{G}^R_{\bmp}(\eps)= \check{G}_\bmp(\ui \veps_n \to \eps+ \ui 0_+)$, $\check{G}^A_{\bmp}(\eps)= \check{G}_\bmp(\ui \veps_n \to \eps- \ui 0_+)$, and the vertex function $\check{\Lambda}^{RA}$ defined on the real-frequency axis satisfies 
\begin{eqnarray}
  \check{\Lambda}^{RA}_{\bmq} (\eps, \omega_{\rm ac}) &=& 
  h_{\rm ac} \hat{\sigma}^z + n_{\rm imp} \int_{\bmp'} \check{V}_{\bmp,\bmp'}
  \check{G}^{R}_{\bmp'} (\eps)  \nonumber \\
  &\times& \check{\Lambda}^{RA}_\bmq (\eps, \omega_{\rm ac}) 
  \check{G}^A_{\bmp'-\bmq}(\eps- \omega_{\rm ac}) \check{V}_{\bmp',\bmp}.
  \label{eq:Lambda03}
\end{eqnarray}

The representation used in Ref.~\cite{Gorkov64}, 
\begin{equation}
  \check{\Lambda}^{RA}_{\bmq} (\eps, \omega_{\rm ac}) =
  \begin{pmatrix} 
    \Lambda^{(1)RA}_\bmq (\eps, \omega_{\rm ac}) \hat{\sigma}^z,
    & \Lambda^{(2)RA}_\bmq (\eps, \omega_{\rm ac})\hat{\sigma}^x \\
    -\Lambda^{(2)RA}_\bmq (\eps, \omega_{\rm ac})\hat{\sigma}^x,
    & \Lambda^{(1)RA}_\bmq (\eps, \omega_{\rm ac})\hat{\sigma}^z 
  \end{pmatrix},
  \label{eq:Lambda_matrix01}
\end{equation}
which is validated to apply even in the present case, transforms Eq.~(\ref{eq:Lambda03}) into a set of linear equations for $\Lambda^{(1)RA}$ and $\Lambda^{(2)RA}$:
\begin{eqnarray}
  \begin{pmatrix}
    \widetilde{\Lambda}^{(1)RA} \\    
    \widetilde{\Lambda}^{(2)RA} 
  \end{pmatrix}
  &=&
  \begin{pmatrix}
    1\\
    0 
  \end{pmatrix}
  +
  \begin{pmatrix}
    {\cal A}, & -{\cal B}\\
    {\cal C}, & {\cal D} 
  \end{pmatrix}
  \begin{pmatrix}
    \widetilde{\Lambda}^{(1)RA} \\    
    \widetilde{\Lambda}^{(2)RA} 
  \end{pmatrix} , 
  \label{eq:Lambda12v02}
\end{eqnarray}
where we have introduced the normalization $\Lambda^{(i)RA}= h_{\rm ac} \widetilde{\Lambda}^{(i)RA}$ ($i=1,2$) for the later convenience. From Eq.~(\ref{eq:Lambda03}), the coefficients from ${\cal A}$ to ${\cal D}$ in the above equation are calculated to be 
\begin{eqnarray}
  {\cal A} &=& \frac{\Gamma_{(-)}}{\pi N(0)} \int_\bmp \left\{ {\cal G}^R_\bmp(\eps) {\cal G}^A_{\bmp_-} (\eps_-)
  + {\cal F}^R_\bmp(\eps){\cal F}^{\dag A}_{\bmp_-} (\eps_- ) \right\}, \label{eq:calA01}\\
  {\cal B} &=& \frac{\Gamma_{(-)}}{\pi N(0)} \int_\bmp \left\{ {\cal G}^R_\bmp(\eps) {\cal F}^{\dag A}_{\bmp_-} (\eps_-)
  + {\cal F}^R_\bmp (\eps){\cal G}^A_{\bmp_-}(\eps_-) \right\}, \label{eq:calB01}\\
  {\cal C} &=& \frac{\Gamma_{(-)}}{\pi N(0)} \int_\bmp \left\{ {\cal G}^R_\bmp(\eps) {\cal F}^A_{\bmp_-} (\eps_-)
  - {\cal F}^R_\bmp(\eps){\cal G}^{\dag A}_{\bmp_-}(\eps_-) \right\}, \label{eq:calC01}\\
  {\cal D} &=& \frac{\Gamma_{(-)}}{\pi N(0)} \int_\bmp \left\{ {\cal G}^R_\bmp(\eps) {\cal G}^{\dag A}_{\bmp_-} (\eps_-)
  - {\cal F}^R_\bmp(\eps){\cal F}^A_{\bmp_-}(\eps_-) \right\}, \label{eq:calD01}
\end{eqnarray}
where $\bmp_- = \bmp- \bmq $ and $\eps_-= \eps- \omega_{\rm ac}$. Here, the scattering rate $\Gamma_{(-)}$ arising from the vertex corrections is given by
\begin{equation}
  \Gamma_{(-)} = \frac{1}{2} \left( \frac{1}{\tau_0}- \frac{1}{3 \tau_{\rm so}} \right), 
  \label{eq:Gamma_-01} 
\end{equation}
where the factor $3$ in front of $\tau_{\rm so}$ should not be forgotten. After integrating over the momentum, we obtain 
\begin{eqnarray}
  {\cal A} &=& \Gamma_{(-)} \frac{|\widetilde{W}_\eps|^2+ |\widetilde{\eps}|^2+ |\widetilde{\Delta}_\eps|^2}
  {|\widetilde{W}_\eps|^2 (2{\rm Im} \widetilde{W}_\eps) } \left( 1- \frac{v_{\rm F}^2 q^2/3}{(2 {\rm Im} \widetilde{W}_\eps)^2}\right), \\
  {\cal B} &=& {\cal C} = \Gamma_{(-)} \frac{\widetilde{\eps} \widetilde{\Delta}_\eps^*
    + \widetilde{\eps}^* \widetilde{\Delta}_\eps}
  {|\widetilde{W}_\eps|^2 (2{\rm Im} \widetilde{W}_\eps) } \left( 1- \frac{v_{\rm F}^2 q^2/3}{(2 {\rm Im} \widetilde{W}_\eps)^2}\right), \\
  {\cal D} &=& \Gamma_{(-)} \frac{|\widetilde{W}_\eps|^2- |\widetilde{\eps}|^2- |\widetilde{\Delta}_\eps|^2}
  {|\widetilde{W}_\eps|^2 (2{\rm Im} \widetilde{W}_\eps) } \left( 1- \frac{v_{\rm F}^2 q^2/3}{(2 {\rm Im} \widetilde{W}_\eps)^2}\right), 
\end{eqnarray}
where $v_F$ is the Fermi velocity, $\widetilde{\eps}= \ui \widetilde{\veps}_n|_{\ui \veps_n \to \eps+ \ui 0_+}$, $\widetilde{\Delta}_\eps= \widetilde{\Delta}|_{\ui \veps_n \to \eps+ \ui 0_+}$, $\widetilde{W}_\eps= \sqrt{\widetilde{\eps}^2- \widetilde{\Delta}_\eps^2}$. Substituting these expressions into Eq.~(\ref{eq:chiQW02}) and comparing the expression of $\delta \check{G}^{R}$ [Eq.~(\ref{eq:delG_RA01})] to the second term on the right hand side of Eq.~(\ref{eq:Lambda03}), the dynamic spin susceptibility is calculated to be 
\begin{eqnarray}
  \frac{1}{\omega_{\rm ac}} {\rm Im} \chi^R_\bmq (\omega_{\rm ac}) &=&
  \frac{N(0)}{4} \int_{-\infty}^\infty \frac{d \eps}{2T}
  \frac{1}{\cosh^2 \left(\frac{\eps}{2T}\right)} \nonumber \\
  && \times
  \left( \frac{\widetilde{\Lambda}_\bmq^{(1)RA}(\eps,0) - 1 }{\Gamma_{(-)}} \right) , 
  \label{eq:chiQW03}
\end{eqnarray}
where $\widetilde{\Lambda}^{(1)RA}$ in the integrand is obtained from Eq.~(\ref{eq:Lambda12v02}), and we have used the small $\omega_{\rm ac}$ limit, $\omega_{\rm ac} \ll \text{min}(T,\Delta)$, which is satisfied above $50$mK for a $10$GHz resonance frequency. Combining the above expression for ${\rm Im}\chi^R_\bmq(\omega_{\rm ac})$ with Eq.~(\ref{eq:Spump01}), we can calculate the additional Gilbert damping constant: 
\begin{eqnarray}
  \delta \alpha &=& \frac{\bra \bra J^2_{\rm sd} \ket \ket}{\hbar^2}
  \frac{N(0)}{4\Gamma_{(-)}} \sum_{\bmq} \int_{-\infty}^\infty \frac{d \eps}{2T}
  \frac{1}{\cosh^2 \left(\frac{\eps}{2T}\right)} \nonumber \\
  && \times
  \frac{{\cal A}(1-{\cal D})-{\cal B}{\cal C}}{(1-{\cal A})(1-{\cal D})+{\cal B}{\cal C}}, 
  \label{eq:Dalpha01}
\end{eqnarray}
which is a manifestation of the spin pumping into superconducting materials.

\section{Results for spin pumping into superconductors\label{Sec:IV}}

In this section, we calculate temperature dependence of spin pumping into superconductors using the formalism developed in the previous section, and show that a pronounced coherence peak appears in the signal immediately below $T_{\rm c}$ even in the presence of the impurity spin-orbit scattering. 

\begin{figure}[t] 
  \begin{center} 
        \scalebox{0.55}[0.55]{\includegraphics{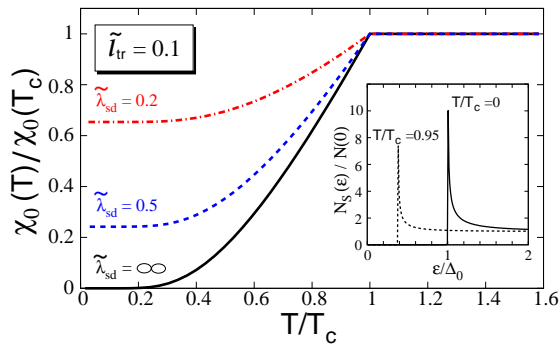}}
  \end{center}
\caption{ 
  (Color online) The uniform susceptibility $\chi_0$ [Eq.~(\ref{eq:GeYoshida01})] as a function of temperature, for a fixed value of $\widetilde{l}_{\rm tr}=0.1$ and for $\widetilde{\lambda}_{\rm sd}= \infty$ (solid line), $0.5$ (dashed line), and $0.2$ (dash-dotted line), where $\widetilde{l}_{\rm tr}$ and $\widetilde{\lambda}_{\rm sd}$ are defined by Eq.~(\ref{eq:ltr-lmd01}). Inset: Density of states [Eq.~(\ref{eq:DOS01})] for $T/T_{\rm c}=0$ and $T/T_{\rm c}=0.95$. 
}
\label{fig_chi0}
\end{figure}

First of all, let us briefly comment on the choice of parameters that characterizes the system under discussion. The present formalism contains two parameters representing the strength of scattering amplitudes: the momentum relaxation time $\tau_{0}$ and the spin-orbit relaxation time $\tau_{\rm so}$. Because it is almost impossible in experiments to tune $\tau_0$ and $\tau_{\rm so}$ independently, it is more convenient to use parameters having a direct connection with physical observables. Thus, we chose to use the following two lengths $l_{\rm tr}$ and $\lambda_{\rm sd}$ to characterize the sample inhomogeneity. The first one, $l_{\rm tr}= v_{\rm F} \tau_{\rm tr}$, is the mean free path which enters into the charge conductivity $\sigma= e^2 N(0) v_{\rm F} l_{\rm tr}/3$, where $e$ is the electron charge and $\tau_{\rm tr}= 1/(\tau_{0}^{-1}+ \tau_{\rm so}^{-1})$ is the transport lifetime. The second one, $\lambda_{\rm sd}= \sqrt{D \tau_{\rm sf}}$, is the spin diffusion length which can be measured experimentally for example by the nonlocal spin valve~\cite{Jedema01}, where $\tau_{\rm sf}= (3/4)\tau_{\rm so}$ is the spin-flip relaxation time~\cite{Fulde68,Niimi13} and $D= v_{\rm F}^2 \tau_{\rm tr}/3$ is the diffusion coefficient. In our numerical calculation, we normalize the above two lengths by the BCS coherence length $\xi_{\rm BCS}= v_{\rm F}/\pi \Delta_0$, where $\Delta_0= 1.76 T_{\rm c}$ is the superconducting gap at $T=0$. Thus, to characterize the system, 
we use the following dimensionless parameters in our numerical calculation: 
\begin{equation}
  \widetilde{l}_{\rm tr} \equiv \frac{l_{\rm tr}}{\xi_{\rm BCS}}, 
  \qquad \text{and} \qquad
  \widetilde{\lambda}_{\rm sd} \equiv \frac{\lambda_{\rm sd}}{\xi_{\rm BCS}}, 
  \label{eq:ltr-lmd01}
\end{equation}
where these two quantities are expressed by $\Gamma_{(+)}$ and $\Gamma_{(-)}$ as
\begin{equation} 
  \widetilde{l}_{\rm tr}= \frac{\pi}{2} \frac{\Delta_0}{\Gamma_{(+)}},
\end{equation}
and
\begin{equation} 
  \widetilde{\lambda}_{\rm sd}
  = \frac{\pi}{\sqrt{12}} \frac{\Delta_0}{\sqrt{\Gamma_{(+)} (\Gamma_{(+)}- \Gamma_{(-)}) }}.
\end{equation}
In the case of Nb, the BCS coherence length is estimated to be of the order of $\xi_{\rm BCS} \approx 200$~nm by using~\cite{Ashcroft-Mermin} $T_{\rm c} \approx 9$~K and $v_{\rm F} \approx 1.4 \times 10^{6}$ ${\rm m}/{\rm s}$. Note that in the dirty limit ($l_{\rm tr} \ll \xi_{\rm BCS}$) which is usually the case in thin-film systems under discussion, the coherence length $\xi$ that is extracted from the upper critical field becomes comparable to the mean free path ($\xi \sim l_{\rm tr}$), and hence $\xi$ differs considerably from the BCS coherence length $\xi_{\rm BCS}$, i.e., $\xi \ll \xi_{\rm BCS}$~\cite{Werthamer-review}.

\begin{figure}[t] 
  \begin{center} 
        \scalebox{0.55}[0.55]{\includegraphics{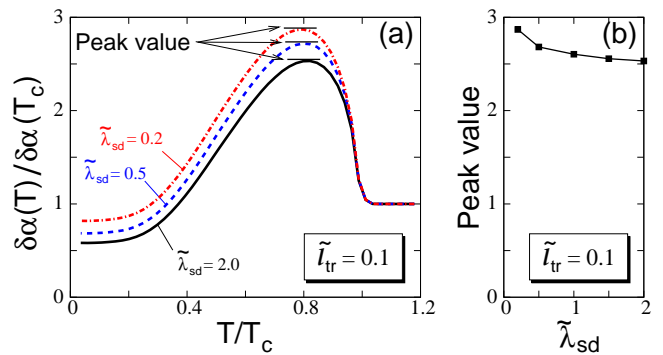}}
  \end{center}
\caption{ 
  (Color online) (a) The additional Gilbert damping constant $\delta \alpha$ [Eq.~(\ref{eq:Dalpha01})] is shown as a function of temperature for a fixed value of $\widetilde{l}_{\rm tr}=0.1$, by varying $\widetilde{\lambda}_{\rm sd}= 2.0$ (solid line), $0.5$ (dashed line), and $0.2$ (dash-dotted line). Here $\widetilde{l}_{\rm tr}$ and $\widetilde{\lambda}_{\rm sd}$ are defined by Eq.~(\ref{eq:ltr-lmd01}). (b) Peak value versus $\widetilde{\lambda}_{\rm sd}$ for $\widetilde{l}_{\rm tr}=0.1$. 
}
\label{fig_Hpp_A}
\end{figure}

Before presenting results for the spin pumping, it is instructive to examine the behavior of the uniform spin susceptibility in order to see the role of the impurity spin-orbit scattering. Considering the fact that a thin-film sample usually has a rather short mean free path $l_{\rm tr}$ in comparison to the BCS coherence length $\xi_{\rm BCS}$, it is realistic to work in a relatively dirty limit $\widetilde{l}_{\rm tr} < 1.0$~\cite{Wakamura14}. The main panel of Fig.~\ref{fig_chi0} shows temperature dependence of the uniform spin susceptibility $\chi_0$, calculated for a fixed value of $\widetilde{l}_{\rm tr}=0.1$ and for several choices of $\widetilde{\lambda}_{\rm sd}$. In the absence of impurity spin-orbit scattering, the curve coincides with the Yoshida function~\cite{Yoshida58} which is suppressed exponentially at low temperatures as $\chi_0 \sim \exp(-\Delta_0/T)$. Upon introducing the impurity spin-orbit scattering and reducing the spin diffusion length $\lambda_{\rm sd}$, there appears a finite susceptibility $\chi_0$ even at the zero temperature limit. In the inset of Fig.~\ref{fig_chi0}, the density of states 
\begin{equation}
  N_S (\eps) = N(0) {\rm Re} \left[ \frac{\widetilde{\eps}}{\widetilde{W}_{\eps}} \right]
  \qquad (\eps > 0), \label{eq:DOS01}
\end{equation}
is plotted as a function of energy $\eps$ for $T/T_{\rm c}= 0$ and $T/T_{\rm c}= 0.95$. As was already mentioned near the end of Sec.~\ref{Sec:II}, the momentum scattering as well as the impurity spin-orbit scattering preserve the time-reversal symmetry of the electron system. Therefore, unlike the case with magnetic impurity scattering, the present system satisfies Anderson's theorem~\cite{Anderson59} and thus a clear superconducting gap appears in the density of states irrespective of the strength of the impurity scattering. This is in stark contrast to gapless superconductors~\cite{Maki-review}, where the time-reversal-symmetry-breaking perturbation destroys the energy gap even when the system is in a superconducting state.

\begin{figure}[t] 
  \begin{center}  
    \scalebox{0.55}[0.55]{\includegraphics{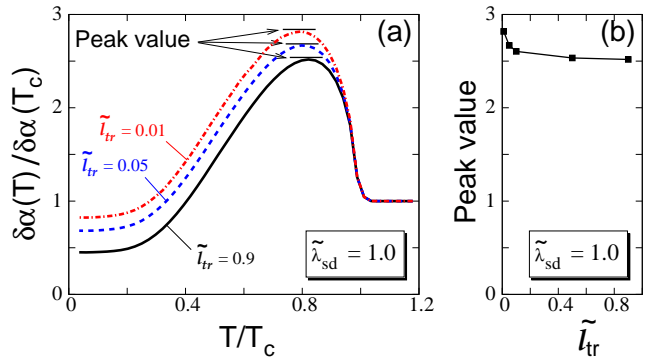}}
  \end{center}
  \caption{
 (Color online) (a) The additional Gilbert damping constant $\delta \alpha$ [Eq.~(\ref{eq:Dalpha01})] is shown as a function of temperature for a fixed value of $\widetilde{\lambda}_{\rm sd}=1.0$, by varying $\widetilde{l}_{\rm tr}= 0.9$ (solid line), $0.5$ (dashed line), and $0.2$ (dash-dotted line). (b) Peak value versus $\widetilde{l}_{\rm tr}$ for $\widetilde{\lambda}_{\rm sd}=1.0$. 
}
\label{fig_Hpp_B}
\end{figure}

Let us now discuss the additional Gilbert damping caused by the spin pumping. Again, we work in a relatively dirty case $\widetilde{l}_{\rm tr} < 1$ and examine the effect of impurity spin-orbit scattering. Figure~\ref{fig_Hpp_A}~(a) shows temperature dependence of the additional Gilbert damping constant, calculated from Eq.~(\ref{eq:Dalpha01}) for a fixed value of $\widetilde{l}_{\rm tr}= 0.1$ and for several values of $\widetilde{\lambda}_{\rm sd}$. First we see that, unlike the ballistic limit calculation ($\widetilde{l}_{\rm tr} = \infty, \widetilde{\lambda}_{\rm sd} = \infty$)~\cite{Hebel-Slichter59}, there remains a finite signal even in the zero temperature limit. This behavior is consistent with that of the uniform spin susceptibility (Fig.~\ref{fig_chi0}), where a nonzero value survives in the $T \to 0$ limit. Second, we find that, even in the presence of the spin-orbit impurity scattering, a clear coherence peak appears immediately below the superconducting transition temperature $T_{\rm c}$. With increasing the impurity spin-orbit scattering and reducing the spin diffusion length ${\lambda}_{\rm sd}$, the height of the coherence peak is gradually enhanced as is seen in the comparison of ``peak value'' there. This tendency is best visible in Fig.~\ref{fig_Hpp_A}~(b), where peak value is plotted as a function of normalized spin diffusion length $\widetilde{\lambda}_{\rm sd}$. Thus, the impurity spin-orbit scattering enhances the height of the coherence peak. We note again that such effects on the coherence peak due to the impurity vertex corrections have not been investigated in the literature. 

A similar conclusion can be derived for the dependence of the coherence peak on the mean free path $l_{\rm tr}$. In Fig.~\ref{fig_Hpp_B}~(a), temperature dependence of the additional Gilbert damping constant is plotted by varying $\widetilde{l}_{\rm tr}$ for a fixed value of $\widetilde{\lambda}_{\rm sd}=1.0$. As one can see in Fig.~\ref{fig_Hpp_B}~(b), again the coherence peak is enhanced by increasing the strength of scattering events, which causes a decrease of the mean free path $l_{\rm tr}$. Therefore, the results displayed in Figs.~\ref{fig_Hpp_A} and \ref{fig_Hpp_B} show that the coherence peak in the additional Gilbert damping constant is enhanced by the momentum scattering as well as the spin-orbit scattering. 

\begin{figure}[t] 
  \begin{center} 
    \scalebox{0.55}[0.55]{\includegraphics{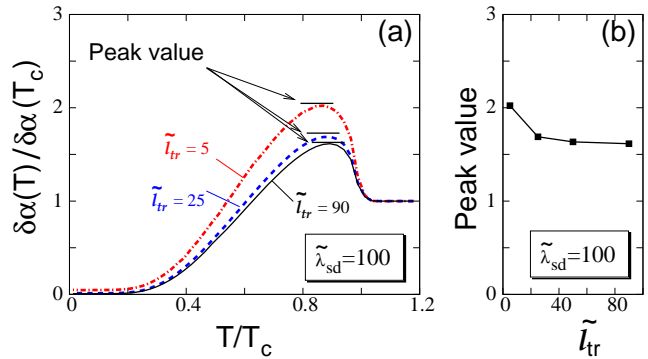}}
  \end{center}
\caption{ 
  (Color online) (a) The additional Gilbert damping constant $\delta \alpha$ [Eq.~(\ref{eq:Dalpha01})] is shown as a function of temperature for a fixed value of $\widetilde{\lambda}_{\rm sd}=100$, by varying $\widetilde{l}_{\rm tr}= 90$ (solid line), $25$ (dashed line), and $5$ (dash-dotted line). (b) Peak value versus $\widetilde{l}_{\rm tr}$ for $\widetilde{\lambda}_{\rm sd}= 100$. 
}
\label{fig_Hpp_C}
\end{figure}

Next, we investigate the case of a weak spin-orbit scattering and hence a very long spin diffusion length, which may be valid when the SS is made of Al~\cite{Bass07}. Figure \ref{fig_Hpp_C}~(a) shows temperature dependence of the additional Gilbert damping constant, calculated for a fixed value of $\widetilde{\lambda}_{\rm sd}= 100$. As in the previous case, we find that there appears a coherence peak below $T_{\rm c}$, and the height of the coherence peak is an increasing function of the strength of the impurity scattering. However, a crucial difference from the previous results with sizable spin-orbit scattering (Figs.~\ref{fig_Hpp_A} and \ref{fig_Hpp_B}) is that the additional Gilbert damping is largely suppressed at low temperatures. This result is in line with the suppression of the uniform spin susceptibility $\chi_0$ at low temperatures (solid curve in Fig.~\ref{fig_chi0}), and also consistent with the ballistic limit calculation~\cite{Hebel-Slichter59}. 

Finally, it is worth mentioning that once we recall a similarity of the quantity under discussion to the spin-lattice relaxation rate in NMR, the above conclusion on the impurity dependence of the coherence peak is consistent with an experimental result of Ref.~\cite{Zheng91}, where the coherence peak measured by the NMR spin-lattice relaxation rate in Cu/Nb multilayers was found to increase with the reduction of the mean free path.

\section{Discussion and conclusion\label{Sec:V}}
The main message of the present paper is that the spin pumping into superconductors can be used as a new technique for probing the spin dynamics in a superconducting {\it thin film}. This is contrasted with the NMR technique used to study a {\it bulk} superconductor. When applied to thin film superconductors, the spin pumping has the following advantages over NMR. First, while NMR suffers from a lowering of signal-to-noise ratio in thin film, the spin pumping does not because it is based on a thin-film technology from the beginning. Second, NMR sometimes requires an isotope substitution to secure probe nuclei, but there is no such issue in the spin pumping as the system is a ferromagnet/target bilayer such that the ferromagnetic resonance condition is ensured by the choice of ferromagnet. Thus, we hope that the present method is applied to a wide range of superconducting materials. 

It would be informative to discuss the difference between the present work and the previous theoretical study~\cite{Morten08}. In short, the main difference lies in the modeling of the ferromagnet/superconductor interface. In the present work, because an {\it insulating} ferromagnet is used as the SI, it is assumed that the exchange interaction $J_{\rm sd}$ at the interface is weak enough, such that it can be dealt with by a perturbative approach (see the last paragraph of Sec.~II in Ref.~\cite{Ohnuma15}). In this case the resultant superconducting gap $\Delta$ as well as the anomalous correlation ${\cal F}(\eps){\cal F}^\dag(\eps)$ survives at the interface, which gives rise to the coherence peak in the spin pumping signal as we have seen in Figs.~\ref{fig_Hpp_A}-\ref{fig_Hpp_C}. In Ref.~\cite{Morten08}, by contrast, because a {\it metallic} ferromagnet is used as the SI (see also Ref.~\cite{Bell08}), it is assumed that the exchange interaction at the interface is so strong that the superconducting gap is completely suppressed there~\cite{comm01}. The latter condition would result in the vanishing of the coherence peak. Note that one data set of Ref.~\cite{Morten08} shows a nonmonotonic behavior of the additional Gilbert damping below $T_{\rm c}$, but such a behavior is visible only in a system containing pair-breaking magnetic impurities not considered in the present work. Under a condition assumed in the present work, i.e., without magnetic impurities, the additional Gilbert damping calculated in Ref.~\cite{Morten08} only shows a monotonic decrease below $T_{\rm c}$. This suggests that the origin of the nonmonotonic behavior found in Ref.~\cite{Morten08} is different from that of the coherence peak found in the present work.

Before conclusion, we add a few remarks on the key points in performing experiments. First, unlike the previous experiment~\cite{Bell08} where a metallic ferromagnet Ni$_{80}$Fe$_{20}$ was used as the spin injector, we consider in the present paper an insulating magnet as the spin injector to simplify the physics involved. Second, it is assumed in our analysis that any spin backflow from the superconducting spin sink is negligibly small. This means that, for a YIG/Nb system, the Nb thickness should be larger than the Nb spin diffusion length ($\lambda_{\rm sd} \sim 50$-$100$~nm for Nb~\cite{Bass07}), while the YIG film should be as thin as possible in order to enhance the spin pumping signal. 

To conclude, we have theoretically studied the spin pumping into superconductors, and predicted that its temperature dependence exhibits a pronounced coherence peak just below $T_{\rm c}$ even in the presence of the impurity spin-orbit scattering. Besides, we have revealed that the height of the coherence peak increases upon the increase of the momentum scattering rate as well as the spin-orbit scattering rate. We propose that the present phenomenon can be used as a new probe for the spin dynamics in a superconducting thin film. Because the present theory fully takes account of the vertex corrections by impurity spin-orbit scattering, it offers a proper description of the diffusive spin dynamics in $s$-wave superconductors. Moreover, since we can draw parallel between the spin pumping signal and the NMR spin-lattice relaxation rate, the present result can also be applied to an analysis of the NMR data when we discuss the effects of the impurity spin-orbit scattering.

\acknowledgments 
We are grateful to S. Onari, K. Tanaka, and K. Matano for valuable discussions, and W. Belzig for useful comments. This work was financially supported by JSPS KAKENHI Grant No. 15K05151. 

\appendix 

\section{Review of the derivation of static susceptibility \label{appendix1}}
In this Appendix, we briefly review the procedure~\cite{Abrikosov62,Gorkov64} to calculate the {\it static} spin susceptibility that fully takes account of the impurity spin-orbit scattering. We consider the uniform and static limit of Eq.~(\ref{eq:delH01}) for the external Hamiltonian: 
\begin{eqnarray}
  \delta {\cal H} &=& 
  - \Big( \sum_\bmp c^\dag_\bmp \hat{\sigma}^z c_\bmp \Big)
  \cdot h_0, 
\end{eqnarray}
where ${\bm h}_0= h_0 \widehat{\bm z}$ is an uniform and static external magnetic field with the Bohr magneton $\mu_{\rm B}$ being absorbed into the definition of $h_0= \mu_{\rm B} H_0$. The uniform spin susceptibility is calculated from the $\bmq \to {\bm 0}$, $\ui \Omega_\nu \to 0$ limit of Eq.~(\ref{eq:chiQW01}):  
\begin{equation}
  \chi_0 = - \frac{\partial}{\partial h_0} \frac{T}{2} \sum_{\veps_n} \int_\bmp \;
      {\rm Tr} \Big[ \hat{\sigma}^z \delta \hat{\cal G}_\bmp (\ui \veps_n) \Big]. 
      \label{eq:chi0v01}
\end{equation}
Here, $\delta \hat{\cal G}_\bmp (\ui \veps_n)$ is the $(1,1)$ component of the matrix Green's function 
\begin{equation}
  \delta \check{G}_\bmp(\ui \veps_n) = \check{G}_\bmp(\ui \veps_n) \check{\Lambda}(\ui \veps_n)
  \check{G}_\bmp(\ui \veps_n), 
  \label{eq:delG01}
\end{equation}
which is proportional to $h_0$. In the above equation, the vertex function $\check{\Lambda}(\ui \veps_n)$ due to the impurity ladder satisfies the following equation,
\begin{eqnarray}
  \check{\Lambda} (\ui \veps_n) &=& h_0 \hat{\sigma}^z + n_{\rm imp} \int_{\bmp'} \check{V}_{\bmp,\bmp'}  \nonumber \\
  &\times&
  \check{G}_{\bmp'} (\ui \veps_n) \check{\Lambda} (\ui \veps_n) \check{G}_{\bmp'}(\ui \veps_n) \check{V}_{\bmp',\bmp}.
  \label{eq:Lambda01}
\end{eqnarray}

The representation similar to Eq.~(\ref{eq:Lambda_matrix01}) transforms Eq.~(\ref{eq:Lambda01}) into a set of linear equations for $\Lambda^{(1)}$ and $\Lambda^{(2)}$: 
\begin{eqnarray}
  \Lambda^{(1)} (\ui \veps_n) &=& h_0 + {A} \Lambda^{(1)}(\ui \veps_n)- {B} \Lambda^{(2)} (\ui \veps_n) , \nonumber \\
  \Lambda^{(2)} (\ui \veps_n) &=& {C} \Lambda^{(1)}(\ui \veps_n) + {D} \Lambda^{(2)} (\ui \veps_n), 
  \label{eq:Lambda12v01}
\end{eqnarray}
where the coeffients $A$, $B$, $C$, and $D$ are expressed as 
\begin{eqnarray}
  {A} &=& \frac{\Gamma_{(-)}}{\pi N(0)} \int_\bmp \left\{ {\cal G}_\bmp(\ui \veps_n) {\cal G}_\bmp (\ui \veps_n)
  + {\cal F}_\bmp(\ui \veps_n){\cal F}^\dag_\bmp(\ui \veps_n) \right\}, \\
  {B} &=& \frac{\Gamma_{(-)}}{\pi N(0)} \int_\bmp \left\{ {\cal G}_\bmp(\ui \veps_n) {\cal F}^\dag_\bmp (\ui \veps_n)
  + {\cal F}_\bmp(\ui \veps_n){\cal G}_\bmp(\ui \veps_n) \right\}, \\
  {C} &=& \frac{\Gamma_{(-)}}{\pi N(0)} \int_\bmp \left\{ {\cal G}_\bmp(\ui \veps_n) {\cal F}_\bmp (\ui \veps_n)
  - {\cal F}_\bmp(\ui \veps_n){\cal G}^\dag_\bmp(\ui \veps_n) \right\}, \\
  {D} &=& \frac{\Gamma_{(-)}}{\pi N(0)} \int_\bmp \left\{ {\cal G}_\bmp(\ui \veps_n) {\cal G}^\dag_\bmp (\ui \veps_n)
  - {\cal F}_\bmp(\ui \veps_n){\cal F}_\bmp(\ui \veps_n) \right\}, 
\end{eqnarray}
and the scattering rate $\Gamma_{(-)}$ is given in Eq.~(\ref{eq:Gamma_-01}). After integrating over the momentum $\bmp$, we obtain ${A}= \Gamma_{(-)} {\widetilde{\Delta}^2}/({\widetilde{\veps}_n^2+ \widetilde{\Delta}^2})^{3/2}$, 
${B}=  {C}= \ui \Gamma_{(-)} {\widetilde{\Delta}\widetilde{\veps}_n}/ ({\widetilde{\veps}_n^2+ \widetilde{\Delta}^2})^{3/2}$, ${D}= \Gamma_{(-)} {\widetilde{\veps}_n^2}/({\widetilde{\veps}_n^2+ \widetilde{\Delta}^2})^{3/2}$.

In order to calculate the susceptibility, it is convenient to use the relation between Eq.~(\ref{eq:delG01}) and the last term of Eq.~(\ref{eq:Lambda01}), which yields
\begin{equation}
  \chi_0 = - \frac{\partial}{\partial h_0} \pi N(0) T \sum_{\veps_n}
  \frac{\Lambda^{(1)} - h_0 }{\Gamma_{(-)}}, 
\end{equation}
where the expression for the summand is transformed into 
\begin{equation}
  \frac{\Lambda^{(1)}-h_0}{\Gamma_{(-)}}
    = \frac{\Delta^2 h_0}{(\veps_n^2+ \Delta^2)}
    \frac{1}{\sqrt{\veps_n^2+\Delta^2}+ \Gamma_{(+)} -\Gamma_{(-)}} 
    \label{eq:delG02}
\end{equation}
after solving Eq.~(\ref{eq:Lambda12v01}). As is seen from the fact that the right-hand side of Eq.~(\ref{eq:delG02}) vanishes in the normal state ($\Delta=0$), it drops the normal state contribution when we firstly integrate over the momentum. Avoiding this singularity by adding the normal state contribution $N(0)$, we finally arrive at Eq.~(\ref{eq:GeYoshida01}).




\end{document}